\documentclass[sigconf]{acmart}


\usepackage{enumitem}
\usepackage[normalem]{ulem}
\usepackage{microtype}
\usepackage{multirow}
\usepackage{booktabs}
\usepackage{float}
\usepackage{graphicx}
\usepackage{tabularx}
\usepackage{wrapfig}
\usepackage{subcaption}
\usepackage{tcolorbox}
\usepackage{xcolor}
\usepackage{booktabs}
\usepackage{fontawesome5} 
\usepackage{enumitem}     
\usepackage{listings}
\usepackage{tcolorbox}
\usepackage{balance}
\tcbuselibrary{skins,breakable,listings}
\AtBeginDocument{%
  }

\copyrightyear{2026}
\acmYear{2026}
\setcopyright{cc}
\setcctype{by}
\acmConference[KDD '26]{Proceedings of the 32nd ACM SIGKDD Conference on Knowledge Discovery and Data Mining V.2}{August 09--13, 2026}{Jeju Island, Republic of Korea}
\acmBooktitle{Proceedings of the 32nd ACM SIGKDD Conference on Knowledge Discovery and Data Mining V.2 (KDD '26), August 09--13, 2026, Jeju Island, Republic of Korea}
\acmDOI{10.1145/3770855.3818858}
\acmISBN{979-8-4007-2259-2/2026/08}
\begin{document}

\title{
PeroMAS: A Multi-agent System of Perovskite Material Discovery}

\author{Yishu Wang}
\authornote{Equal contribution.}
\affiliation{%
  \institution{Southeast University}
  \city{Nanjing}
  \country{China}}
\email{yishu1031@seu.edu.cn}

\author{Wei Liu}
\authornotemark[1]
\affiliation{%
  \institution{Southeast University}
  \city{Nanjing}
  \country{China}}
\email{wei.liu@seu.edu.cn}

\author{Yifan Li}
\affiliation{%
  \institution{Southeast University}
  \city{Nanjing}
  \country{China}}
\email{230268413@seu.edu.cn}

\author{Shengxiang Xu}
\affiliation{%
  \institution{Southeast University}
  \city{Nanjing}
  \country{China}}
\email{xushx@seu.edu.cn}

\author{Xujie Yuan}
\affiliation{%
  \institution{Sun Yat-Sen University}
  \city{Zhuhai}
  \country{China}}
\email{yuanxj8@mail2.sysu.edu.cn}

\author{Ran Li}
\affiliation{%
  \institution{The Hong Kong University of Science and Technology}
  \city{Hong Kong}
  \country{China}}
\email{rlibb@connect.ust.hk}

\author{Yuyu Luo}
\affiliation{%
  \institution{The Hong Kong University of Science and Technology (Guangzhou)}
  \city{Guangzhou}
  \country{China}}
\email{yuyuluo@hkust-gz.edu.cn}

\author{Jia Zhu}
\affiliation{%
  \institution{Zhejiang Normal University}
  \city{Jinhua}
  \country{China}}
\email{jiazhu@zjnu.edu.cn}

\author{Shimin Di}
\authornote{Corresponding author.}
\affiliation{%
  \institution{Southeast University}
  \city{Nanjing}
  \country{China}}
\email{shimin.di@seu.edu.cn}

\author{Min-Ling Zhang}
\authornotemark[2]
\affiliation{%
  \institution{Southeast University}
  \city{Nanjing}
  \country{China}}
\email{zhangml@seu.edu.cn}

\author{Guixiang Li}
\authornotemark[2]
\affiliation{%
  \institution{Southeast University}
  \city{Nanjing}
  \country{China}}
\email{guixiang.li@seu.edu.cn}

\renewcommand{\shortauthors}{Yishu Wang et al.}

\begin{abstract}
As a pioneer of the third-generation photovoltaic revolution, Perovskite Solar Cells (PSCs) are renowned for their superior optoelectronic performance and cost potential.
The development process of PSCs is precise and complex, involving a series of closed-loop workflows such as literature retrieval, data integration, experimental design, and synthesis.
However, existing AI perovskite approaches focus predominantly on discrete models, 
including material design, process optimization, and property prediction.
These models fail to propagate physical constraints across the workflow, hindering end-to-end optimization.
In this paper,
we propose a multi-agent system for perovskite material discovery, named PeroMAS.
We first encapsulated a series of perovskite-specific tools into Model Context Protocols (MCPs). 
By planning and invoking these tools, PeroMAS can design perovskite materials under multi-objective constraints, covering the entire process from literature retrieval and data extraction to property prediction and mechanism analysis.
Furthermore, we construct an evaluation benchmark by perovskite human experts to assess this multi-agent system.
Results demonstrate that, compared to single Large Language Model (LLM) or traditional search strategies, our system enhances discovery efficiency.
It iteratively tracks and diagnoses candidate proposals under multi-objective constraints, refining toward target objectives through closed-loop strategy adaptation.
Notably, we verify PeroMAS's effectiveness in the physical world through real synthesis experiments.
\end{abstract}


\maketitle

\newcommand\kddavailabilityurl{https://github.com/DEFENSE-SEU/PeroMAS}
\ifdefempty{\kddavailabilityurl}{}{
\begingroup\small\noindent\raggedright\textbf{Resource Availability:}\\
The source code and datasets of this paper have been made publicly available at \url{\kddavailabilityurl}.
\endgroup
}

\section{Introduction}

Global climate change and traditional energy crises have established the development of renewable energy as a central objective for sustainable development~\citep{lee2025meeting}.
Given that solar energy is the most abundant and accessible renewable resource, solar cells serve as a critical pillar for achieving carbon neutrality and ensuring energy security~\citep{haegel2017terawatt}. 
A primary focus of photovoltaic technology is improving the power conversion efficiency (PCE), which measures the conversion of incident light power into output electrical power. Crystalline silicon materials, characterized by stable systems and mature device processing, currently support large-scale energy supplies with a PCE nearing 26\%. 
However, the Shockley-Queisser limit restricts single-junction silicon solar cells to a theoretical maximum PCE of approximately 33\%~\citep{wang2023carrier}, leading to diminishing marginal returns in silicon-based research. 

In contrast, 
perovskite solar cells (PSCs)~\citep{li2025stabilizing, wu2025resilience} offer high manufacturability and 
rapid performance improvements.
Over the past decade, the PCE of PSCs has increased from an initial 3.8\% to over 26\%, exceeding 33\% in tandem configurations. Specifically, tandem PSCs utilizing perovskite materials can achieve efficiencies as high as 45\%, successfully exceeding the fundamental limits of crystalline silicon devices~\citep{zhao2024suppressed, wu2025resilience}. Despite these advantages, the large-scale application of PSCs remains hindered by severe challenges, including limited long-term stability~\citep{li2023highly, zhang2025defect, sang2025improving}, the toxicity of lead-based materials~\cite{li2024biotoxicity, yang2023reducing}, and poor reproducibility in scalable manufacturing~\cite{weerasinghe2024first}. 
Overcoming these challenges requires optimization of the perovskite material discovery process, 
including material compositions, lattice structure, and processing parameters~\citep{peng2025halide}.

Perovskite material discovery constitutes a multi-step scientific workflow defined by four core tasks: knowledge preparation, design, dry-lab simulation, and wet-lab validation~\citep{wu2024universal}. 
Knowledge preparation serves as the research foundation, comprising literature retrieval, the aggregation of historical data, and the extraction of dispersed experimental parameters—such as A/B/X site occupancy, precursor concentration, spin-coating speed, and annealing temperature. These parameters, along with performance metrics (e.g., PCE, $V_{oc}$, $J_{sc}$, and FF), are organized into structured databases~\citep{jacobsson2022open}. Based on knowledge, researchers initiate material design by 
generating verifiable hypotheses~\citep{cai2022discovery, zhu2026survey}. This process involves 
composition design~\citep{jeon2015compositional}, 
structural design~\citep{chen2020strain}, and 
process design~\citep{jeon2014solvent} to formulate 
executable candidate solutions~\citep{nikolaev2016autonomy, wang2025data}. 
To minimize prohibitive trial-and-error costs, dry-lab simulations are essential~\citep{tao2021machine}. Specifically, density functional theory (DFT) calculations or machine learning (ML) models are employed to predict band structures, formation energies, and optoelectronic properties, enabling feasibility screening and providing necessary mechanistic analysis~\citep{tao2019absolute, lu2018accelerated}. Finally, wet-lab experiments represent the physical realization phase, encompassing synthesis, thin-film fabrication, device assembly, and rigorous performance testing and validation~\citep{zhang2024toward}.

\begin{figure}[!t]
\centering
\includegraphics[width=\linewidth]{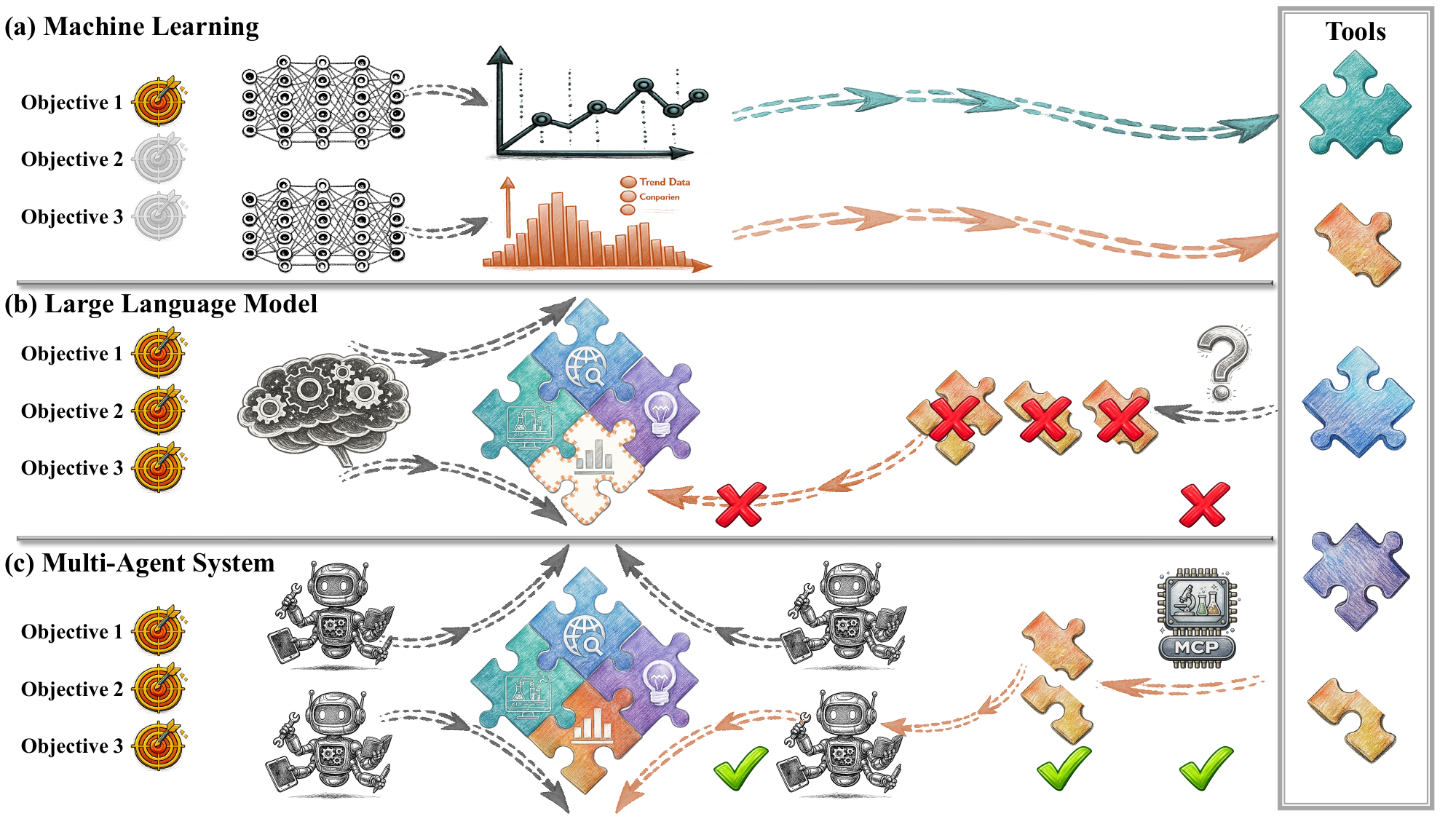}
\vspace{-10px}
\caption{
(a) Although ML methods focus on single-objective mapping, they create high-precision specialized models that populate the PSC tool pool; 
(b) LLMs integrate domain knowledge for multi-objective tasks but may suffer from uncontrolled constraints and lack invocation for PSC tools;
(c) We propose PeroMAS to orchestrate functional agents that dynamically retrieve and utilize tools from the PSC tool pool, iteratively refining solutions under multi-objective constraints.}
\vspace{-12px}
\label{fig-ai3paradigm}
\end{figure}

However, the discovery of novel perovskite materials using traditional trial-and-error methods remains inefficient. 
As shown in Figure~\ref{fig-ai3paradigm},
three primary artificial intelligence (AI) paradigms have been introduced to facilitate research on perovskite materials.
\begin{itemize}[leftmargin=*]
    \item \textbf{Machine Learning (ML)}: Utilizing ML models such as Random Forest (RF)~\citep{zhou2019deep}, Support Vector Machines (SVM)~\citep{gonen2011multiple}, and Graph Neural Networks (GNN)~\citep{wu2020comprehensive} to establish mappings between material structures or processing parameters and 
    device performance.
    Leveraging the powerful fitting capability~\citep{chabra2020deep}, ML models enable the rapid prediction of properties, e.g., PCE, stability, and bandgap. 
    For example, the GNoME project successfully predicted millions of stable crystal structures, including perovskites~\citep{merchant2023scaling}. Nevertheless, this paradigm generally supports only the automation of specific tasks, e.g., property prediction models fail to inform subsequent composition design directly. 
    Furthermore, ML-based perovskite studies predominantly focus on single-objective optimization, typically maximizing PCE, rather than on multi-objective optimization~\citep{macleod2022self, xu6079664machine}.

    \item \textbf{Large Language Model (LLM)}: This paradigm involves pre-training or fine-tuning LLMs on perovskite literature to embed domain-specific knowledge. 
    By leveraging advanced semantic understanding and generation capabilities, LLMs automate knowledge preparation, assisting diverse research tasks. 
    For instance, Perovskite-LLM fine-tunes on 2,214 papers to support knowledge retrieval, literature reviews, experimental design, and complex problem-solving~\citep{liu2025perovskite}. 
    Similarly, Perovskite-R1 
    identifies high-quality small-molecule additives that are subsequently validated through wet-lab experiments~\citep{wang2025perovskite}. 
    Despite these advances, generative AI often underperforms on fundamental and specialized sub-tasks~\citep{schick2023toolformer}.
    Especially, these approaches primarily internalize knowledge into the model, failing to effectively leverage stable external tools to further extend their capabilities.

    \item \textbf{Multi-agents}: Built upon LLMs, multi-agent systems incorporate capabilities for planning, tool-use, reflection, and environmental interaction~\citep{zhang2025exploring}.
    When integrated with self-driving laboratories, agents enable autonomous synthesis and characterization~\citep{boiko2023autonomous}. 
    Their flexible task orchestration and global planning potential offer a pathway to resolve model fragmentation across the research workflow. 
    For example, focusing exclusively on the single task of composition design, \citet{lee2025enhanced} proposed a multi-agent framework—comprising formula proposal, condition extraction, text gradient, and structure formatting agents—to discover novel double perovskites. 
    However, this paradigm within the perovskite field remains insufficient, and a dedicated agent system covering the full process has not yet been developed. 
\end{itemize}

As summarized in Figure~\ref{fig-ai3paradigm},
existing methods either automate isolated tasks without a unified workflow, or, in the generative setting, lack closed-loop tracking and coordination of perovskite-specific constraints across the design--simulation--analysis stages.
Furthermore, most methods only focus on a single perovskite metric while ignoring other metrics, for example, optimizing PCE alone may compromise stability~\citep{xu2025multi}.

To address above challenges, this paper presents the first specialized multi-agent system for perovskite material discovery.
As illustrated in Figure~\ref{fig-framework}, the system utilizes a hierarchical architecture with a 
\textit{meta agent} and multiple \textit{functional agents}.
Meta agent is responsible for global task orchestration, constraint instantiation, and decision-making, ensuring workflow continuity across multiple tasks. To enhance task-solving capabilities, the system adopts a task-oriented approach, utilizing the Model Context Protocol (MCP)~\citep{hou2025model} to equip functional agents with specialized tools, including 
\textit{miner agent},
\textit{designer agent},
\textit{emulator agent},
and
\textit{analyst agent}.
Miner agent focuses on knowledge preparation, automating literature retrieval and data extraction to establish data priors as foundational hypotheses. 
Designer agent executes combinatorial design for compositions, structures, and processes, translating abstract hypotheses into concrete experimental parameters. 
Emulator agent is designed for dry-lab simulations where prediction models are automatically invoked to perform feasibility screening and multi-property forecasting. 
Analyst agent tracks constraint gaps, diagnoses limiting factors, and recommends strategy adjustments to close the feedback loop for subsequent iterations. The primary contributions of this paper are as follows:
\begin{itemize}[leftmargin=*]
\item 
We construct the first full-process multi-agent system in the perovskite domain that enables closed-loop iteration across knowledge preparation, design, and dry-lab simulation, addressing the lack of an end-to-end agentic workflow in this subdomain.
\item 
By orchestrating functional agents via a Meta Agent and standardizing the integration of heterogeneous tools via the unified MCP interface, the system iteratively tracks and refines candidate solutions under multi-objective constraints. 
\item 
We validate the system through a wet-lab experiment. Under efficiency-oriented and low-toxicity constraints, PeroMAS iteratively refined candidate proposals toward target objectives, yielding a lead-reduced (50\%) recipe with a measured PCE of 17\%, verifying the system's practical viability in physical synthesis.
\item 
We construct the first evaluation benchmark by perovskite human experts to assess PeroMAS.
It includes an atomic-level dataset for subtasks and a full-task dataset containing complex multi-objective queries to facilitate research on agent planning and coordination.
\end{itemize}

\begin{figure*}[!t]
    \centering
    \includegraphics[width=0.95\textwidth]{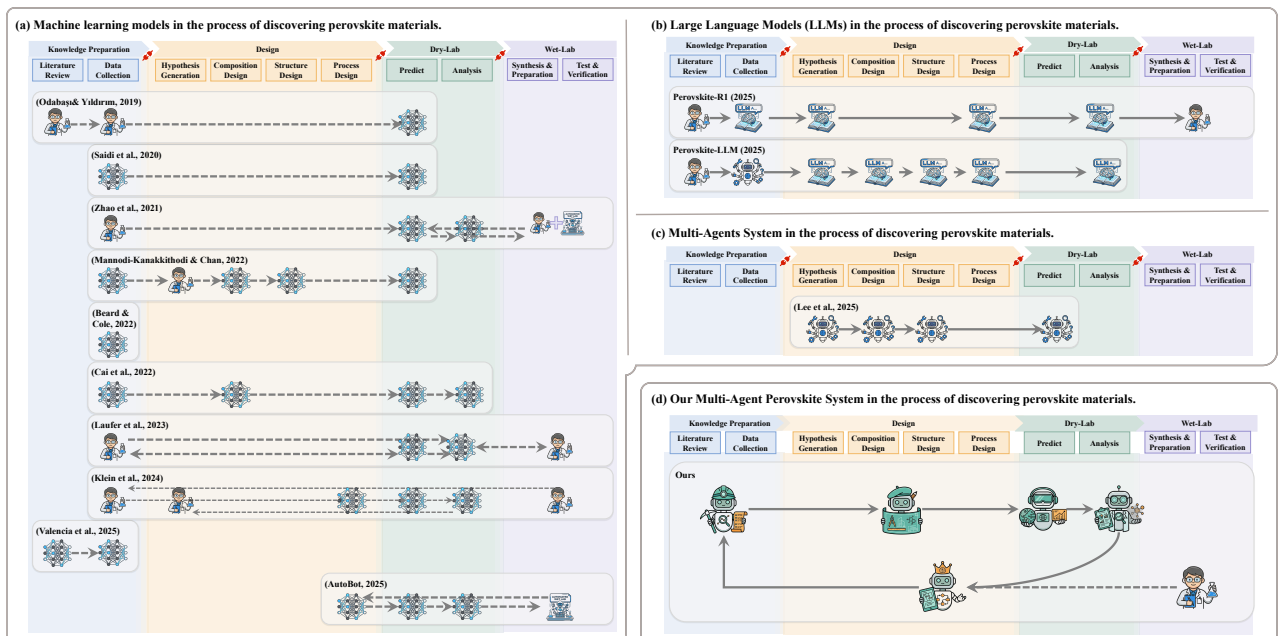}
    \vspace{-10px}
    \caption{A task-oriented overview of related work on AI for PSCs, including knowledge preparation, design, dry-lab, and wet-lab.
    PeroMAS automatically executes the full process for perovskite research and validates the candidates designed by PeroMAS through wet-lab experiments.}
    \vspace{-15px}
    \label{fig-related-work}
\end{figure*}

\section{Related Work}
In the era of data-intensive science, AI not only accelerates research through large-scale processing and automation but also expands the boundaries of scientific discovery by extracting patterns and knowledge from complex datasets~\citep{zhu2025survey}. 
This section first reviews the progress of MAS across various scientific domains. 
Subsequently, we organize existing literature according to the core tasks of perovskite research.

\subsection{Scientific Multi-Agent Systems}
Early scientific agents such as ChemCrow, ChatInvent and LLM-based inorganic synthesis predictors have shown that tool-augment\-ed LLMs can help chemistry, drug discovery and earth observation~\citep{bran2023chemcrow, he2026democratising, zhao2025comprehensive, yao2026remoteagent}. However, these single-agent or task-specific systems remain limited in coordinating multi-stage scientific workflows with persistent planning and iterative feedback. Multi-agent systems (MAS) further decompose complex scientific objectives into executable workflows in the digital realm~\citep{panapitiya2025autolabs, xu2025robustflow}. In the physical realm, MAS can be coupled with robotic platforms to autonomously execute experiments. Together, these capabilities can reduce the burden on human researchers.

Coscientist represents a milestone in automated scientific research. This system develops a Planner responsible for high-level reasoning, task decomposition, and the scheduling of Web Searcher, Docs Searcher, Code Execution, and Automation Interface. It successfully accelerates chemical research across six tasks~\citep{boiko2023autonomous}. Furthermore, emerging research continuously facilitates automated research within the chemical domain. ChemAgents employs a task manager to coordinate literature mining, computational modeling, and robotic operations~\citep{song2025multiagent}, while AutoLabs utilizes a graph-based architecture and self-check mechanisms to convert natural language instructions into validated, near-expert-level protocols for high-throughput liquid handlers~\citep{panapitiya2025autolabs}.

In the biomedical domain, The Virtual Lab simulates a collaborative research team using hierarchical specialized agents and human feedback to reinforce scientific reasoning~\citep{swanson2025virtual}. Similarly, CRISPR-GPT automates the gene-editing workflow validated by the Gene-editing Bench~\citep{qu2025crispr}. Furthermore, STELLA features a self-evolving mechanism that refines decision-making strategies to address novel biomedical challenges~\citep{jin2025stella}.

In the materials domain, SciAgents integrates large-scale ontological knowledge graphs with multi-agent systems, realizing exploration capabilities that transcend human limits~\citep{ghafarollahi2025sciagents}. AtomAgents further integrates physics-aware reasoning with molecular simulations to address complex alloy design tasks through a collaborative multi-agent strategy~\citep{ghafarollahi2024atomagents}.

MAS have similarly demonstrated potential in other scientific domains. PharmAgents automates drug discovery workflows~\citep{gao2025pharmagents}, while Foam-Agent automates end-to-end computational fluid dynamics workflows through LLM-based multi-agent coordination, MCP-based tool integration, and dependency-aware scheduling~\citep{pan2026automating}. Although these works validate the utility of MAS in various fields, the discovery of PSC still lacks a domain-specific MAS that coordinates the preparation, design, and dry-lab screening of knowledge and validation of wet-lab under coupled material constraints.

\subsection{Artificial Intelligence for Perovskite}
\label{sec-Realted-ai4PSC}
To accelerate perovskite research, artificial intelligence (AI) has been extensively explored across its multi-step workflow.

\textbf{Knowledge Preparation.} As the foundation of data-driven discovery, this stage faces the dual challenges of data scarcity and literature overload. While manual efforts like the Perovskite Database~\citep{jacobsson2022open} provide structured high-quality data, they remain labor-intensive and prone to cognitive bias. Conversely, automated approaches, ranging from DFT calculations~\citep{beard2022perovskite} and high-throughput experimentation~\citep{zhao2021discovery} to LLM-based extraction~\citep{liu2025perovskite}, offer scalability but often suffer from model obsolescence and irrelevant literature that can mislead task-specific decision-making.

\textbf{Prediction and Screening.} By learning the mapping between different material designs and their properties, machine learning (ML) models have successfully predicted key properties such as bandgaps, stability, and J-V characteristics~\citep{saidi2020machine, hartono2023stability}. These discriminative and predictive capabilities enable the rapid screening of vast chemical spaces, narrowing down thousands of candidates to a few promising candidates~\citep{ma2025accelerating, cai2022discovery}. However, these models are primarily confined to attempting different combinations of known perovskite components, device structures, and processing conditions, rather than coordinating constraints and feedback across downstream design stages.

\textbf{Analysis and Design.} To transcend simple property mapping, recent research integrates mechanistic analysis and generative design. While early ML and explainable AI methods identified statistical associations between fabrication parameters and performance~\citep{klein2024discovering}, lacking deep reasoning capabilities. Perovskite-LLM and Perovskite-R1 address this problem by synthesizing literature insights for knowledge retrieval and reasoning~\citep{liu2025perovskite, wang2025perovskite}. Furthermore, ~\citet{lee2025enhanced} utilize a surrogate-assisted agent framework to optimize double perovskite compositions. Nevertheless, these systems currently focus on single-task optimization, typically composition or processing, thereby neglecting the multi-objective interplay among composition, structure, and processing~\citep{correa2017promises, oh2025algorithm}.

\textbf{Wet Experimentation.} Finally, robotic platforms like RAPID and AURORA have demonstrated the potential for high-throughput synthesis and characterization, significantly compressing experimental cycles~\citep{li2019robot, macleod2020self}. However, fully autonomous discovery remains hindered by the heterogeneity of physical equipment and the inherent stochasticity of physical experiments, which sever the seamless interaction between digital and physical realms ~\citep{mehr2020universal, li2023real, canty2025science}.

\section{Methodology}
\label{sec-method}

This section presents our multi-agent framework for the autonomou\-s discovery of perovskite materials. We first formalize the discovery task as a multi-objective optimization problem and outline the system architecture in Section~\ref{sec-method-overview}. Section~\ref{sec-method-meta-agent} details the Meta Agent, the system's central orchestrator. Subsequently, Section~\ref{sec-method-function-agent} describes the four functional agents (Miner, Designer, Emulator, and Analyst), focusing on their tool execution via the Model Context Protocol (MCP)~\citep{hou2025model}. Finally, Section~\ref{sec-method-wetlab} validates the system through a standardized wet-lab synthesis procedure.

\begin{figure*}[htbp]
    \centering
    \includegraphics[width=0.9\textwidth]{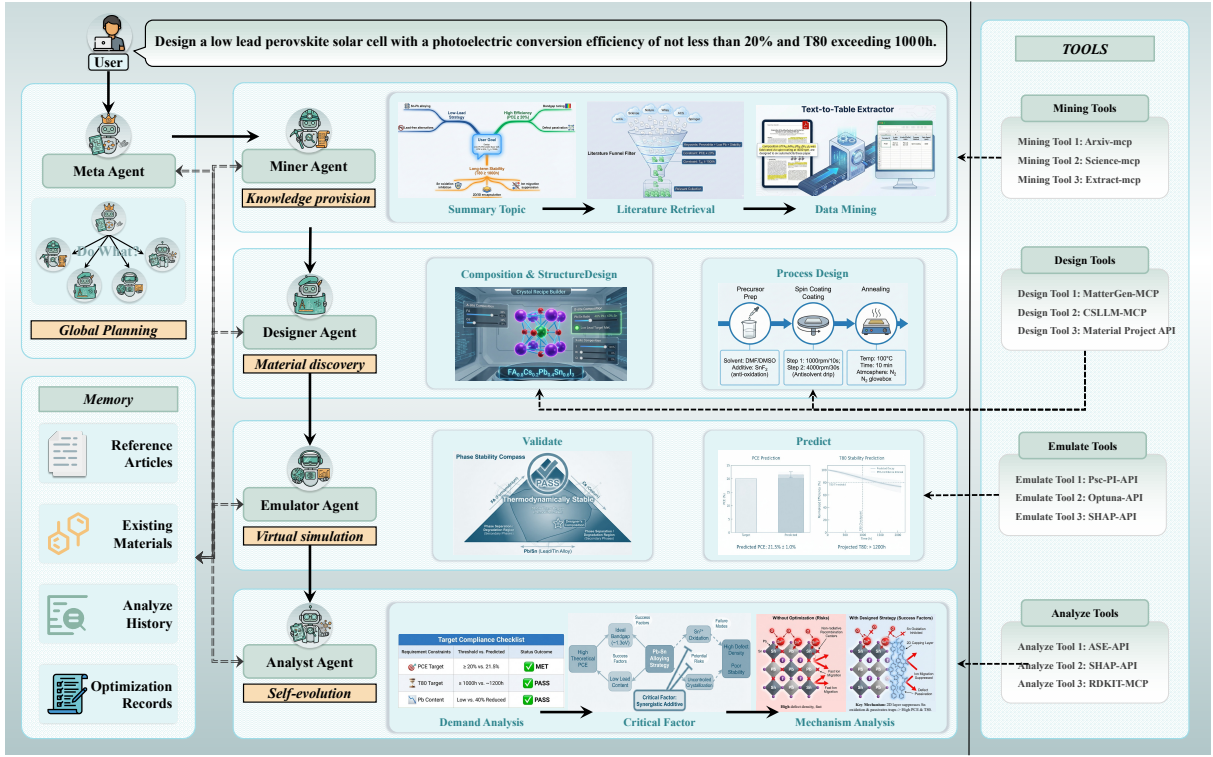}
    \vspace{-10px}
    \caption{The system is centrally orchestrated by a \textbf{Meta Agent} responsible for global planning and memory management. It directs four functional agents, which are equipped with specialized domain tools via MCP to execute the closed-loop discovery workflow—spanning knowledge preparation, inverse design, property prediction, and mechanistic analysis.}
    \vspace{-10px}
    \label{fig-framework}
\end{figure*}

\subsection{Overview}\label{sec-method-overview}

Perovskite discovery faces two fundamental challenges: high-dimen\-sional parameter coupling and conflicting multi-objective trade-offs (e.g., efficiency vs. stability)~\citep{merchant2023scaling}. To address these, we formulate the discovery process as a dynamic optimization problem. Unlike existing machine learning approaches confined to isolated tasks~\citep{low2022mapping,xu2025multi}, we propose PeroMAS (Figure~\ref{fig-framework}), a task-flow-oriented multi-agent system. It adopts a hierarchical architecture: a Meta Agent orchestrates central planning, while four functional agents (Miner, Designer, Emulator, Analyst) execute specialized subtasks using the Model Context Protocol (MCP) and shared memory, effectively resolving the fragmentation of traditional pipelines.
\subsection{Meta Agent}\label{sec-method-meta-agent}

The Meta Agent serves as the system's central reasoning unit, responsible for high-level task orchestration and global decision-making. Distinct from functional agents that rely on external tools, the Meta Agent operates as a logic-driven planner, translating user-defined objectives into an executable PDCA (Plan-Do-Check-Act) workflow. At the $t$-th iteration, it processes a composite state vector:
\begin{equation}
    S_t = \{ \mathcal{G}, \mathcal{M}, \mathcal{R}_{t-1} \},
\end{equation}
where $\mathcal{G} = \langle \mathcal{O}_{\text{obj}}, \mathcal{C}_{\text{const}} \rangle$ denotes the user-defined goal, with $\mathcal{O}_{\text{obj}}=\{f_i(x)\}_{i=1}^{n}$ representing measurable target objectives (e.g., PCE, $T_{80}$, toxicity, and stability) and $\mathcal{C}_{\text{const}}=\{g_j(x)\}_{j=1}^{m}$ representing computable physical constraints (e.g., charge balance, stoichiometry, ABX$_3$ validity, and tolerance-factor validity); $\mathcal{M}$ denotes the historical memory storing decision trajectories and execution logs; and $\mathcal{R}_{t-1}$ represents diagnostic feedback from the previous cycle.

Leveraging the Chain-of-Thought (CoT) mechanism~\citep{wei2022chain}, the Meta Agent executes a three-stage reasoning process:
\begin{enumerate}[leftmargin=1.2em, itemsep=0pt, topsep=2pt, parsep=0pt, label=(\arabic*)]
    \item \textbf{Goal Alignment:} It evaluates whether the feedback $\mathcal{R}_{t-1}$ meets the termination criteria defined in $\mathcal{G}$. If targets are achieved or iteration limits reached, the process concludes.
    \item \textbf{Reflection and Attribution:} For unfulfilled objectives, the agent analyzes $\mathcal{M}$ to identify root causes of previous failures, ensuring the avoidance of proven ineffective paths.
    \item \textbf{Strategy Formulation:} Based on the analysis, it instantiates abstract constraints into concrete scientific hypotheses $\mathcal{H}_t$ and formulates the exploration strategy for the next iteration.
\end{enumerate}

The resulting orchestration instruction set $\mathcal{P}_t$ is:
\begin{equation}
    \mathcal{P}_t = \{ \mathcal{I}_{\text{Miner}}, \mathcal{I}_{\text{Designer}}, \mathcal{I}_{\text{Emulator}}, \mathcal{I}_{\text{Analyst}} \}.
\end{equation}

These instructions define the contextual dependencies within the workflow. To optimize computational efficiency, the Meta Agent employs an adaptive pruning mechanism: by referencing $\mathcal{M}$, it identifies redundant sub-tasks (e.g., where valid data already exists) and sets their corresponding instructions to $\emptyset$ (SKIP), thereby concentrating computational resources on unexplored design spaces.

\subsection{Function Agents}\label{sec-method-function-agent}

Guided by the Model Context Protocol (MCP)~\citep{hou2025model}, four functional agents—Miner, Designer, Emulator, and Analyst—execute the closed-loop R\&D process. These modules map to knowledge, design, prediction, and analysis phases, respectively. By ingesting upstream data and planning instructions, they ensure lossless context transmission while autonomously leveraging domain tools to generate concrete experimental results.

\subsubsection{Miner Agent}\label{sec-method-func-miner-agent}

Miner Agent is designed to address the ``cold-start'' challenge in the high-dimensional search space by constructing a domain knowledge base $\mathcal{D}_{\text{obs}}$ from unstructured literature.

\textbf{Mechanism and Tool Integration:} The agent executes a two-stage ``Retrieval-Extraction'' workflow.
First, it synthesizes the global goal $\mathcal{G}$ and the current instruction $\mathcal{I}_{\text{Miner}}$ into a retrieval query set $\mathcal{Q}$. By invoking the Literature Retrieval Interface (encapsulating APIs such as arXiv and ScienceDirect via MCP), it acquires a collection of highly relevant documents $\mathcal{D}_{\text{doc}}$.
Subsequently, the agent utilizes a Semantic Extraction Toolchain—which synergizes domain-specific extraction models with LLMs~\citep{wan2024tokens, beltagy2019scibert}—to construct an extraction function $f_{\text{ext}}$. This function parses unstructured texts $d \in \mathcal{D}_{\text{doc}}$ into hybrid knowledge entries $x_k$ containing design parameters ($\mathcal{X}_{\mathit{design}}$), performance metrics ($\mathcal{Y}_{\mathit{perf}}$), and mechanistic insights ($\mathcal{S}_{\mathit{mech}}$):
\begin{equation}
    x_k = f_{\mathit{ext}}(d \mid \mathcal{G}, \mathcal{I}_{\text{Miner}}) = \langle \mathcal{X}_{\mathit{design}}, \mathcal{Y}_{\mathit{perf}}, \mathcal{S}_{\mathit{mech}} \rangle.
\end{equation}

Finally, the output dataset $\mathcal{D}_{\text{obs}} = \{ x_k \}_{k=1}^{N}$ is automatically merged into the system's historical memory $\mathcal{M}$, providing the data foundation for downstream tasks.

\subsubsection{Designer Agent}\label{sec-method-func-designer-agent}

Designer Agent addresses the high-dimensional coupling challenge of composition, structure, and process ($c, s, p$) by performing inverse design within the combinatorial space.

\textbf{Mechanism and Tool Integration:} The agent employs a context-aware ``Joint Generation-Constraint Filtering'' mechanism driven by a ``Generation-Reasoning'' dual-engine toolchain. First, conditioning on the instruction $\mathcal{I}_{\text{Design}}$ and prior data $\mathcal{D}_{\text{obs}}$, it invokes the Material Generation Interface (integrating fine-tuned deep generative models~\citep{zeni2025generative}) to efficiently sample candidate compositions ($c$) and structures ($s$) from the latent space. Subsequently, a Synthesis-Aware LLM~\citep{song2025accurate} complements these candidates by determining specific synthesis parameters ($p$) and acting as a discriminator to filter out physically unfeasible proposals based on hard constraints $g(x)$. The joint generation process is formalized as:
\begin{equation}
    x_{\mathit{cand}} = f_{\mathit{gen}}(\mathcal{I}_{\mathit{Design}} \mid \mathcal{D}_{\mathit{obs}}) = \langle c, s, p \rangle.
\end{equation}

The final output is a constraint-verified experimental proposal set $\boldsymbol{X}_{\mathit{new}} = \{ x_i \}_{i=1}^{K}$, 
then fed into the virtual screening phase.

\subsubsection{Emulator Agent}\label{sec-method-func-emulator-agent}

Emulator Agent functions as a high-throughput virtual screening funnel to address the expensive, non-analytical black-box objective $F(x)$ defined in the Overview.

\textbf{Mechanism and Tool Integration:} The agent employs a multi-dimensional property prediction workflow powered by a heterogeneous prediction toolchain. Given the candidate set $\mathcal{X}_{\text{new}}$ from the Designer Agent and prior data $\mathcal{D}_{\text{obs}}$, it invokes two core modules in parallel: the Graph Structure Prediction Interface, which mounts Graph Neural Networks~\citep{xie2018crystal} to predict intrinsic material properties, and the Machine Learning Regression Engine, which integrates classic algorithms~\citep{hu2024interpretable} to infer device-level metrics. For each candidate $x$, the black-box objective is instantiated as
\begin{equation}
    F(x)=\hat{\mathbf{y}}(x)
    = \{f_{\text{PCE}}(x), f_{T_{80}}(x), f_{\text{Stability}}(x), f_{\text{Toxicity}}(x)\},
\end{equation}
where each $f_i(\cdot)$ is a prediction function for the $i$-th objective in $\mathcal{O}_{\text{obj}}$. The uncertainty of each prediction is estimated as
\begin{equation}
    \sigma_i(x)=\mathrm{Std}[\hat{y}_i(x)],
\end{equation}
which represents the prediction standard deviation. The utility score used for candidate ranking is computed by
\begin{equation}
    U(x)=\sum_{i=1}^{n} w_i s_i(x),
    \quad s_i(x)\in[0,1],
\end{equation}
where $s_i(x)$ denotes the satisfaction score of the $i$-th objective and $w_i$ is its corresponding weight. The result is an evaluated candidate set $\mathcal{X}_{\text{eval}}$, from which high-potential candidates are selected according to $U(x)$.

\subsubsection{Analyst Agent}\label{sec-method-func-analyst-agent}

The Analyst Agent serves as the pivotal hub for the ``Check'' phase, synthesizing multidimensional data to diagnose performance discrepancies and close the discovery loop.

\textbf{Mechanism and Tool Integration:} The agent executes a two-layer ``Rule Filtering-Attribution'' pipeline. First, it aggregates the full data stream ($\mathcal{D}_{\mathit{obs}}, \boldsymbol{X}_{\mathit{new}}, \boldsymbol{X}_{\mathit{eval}}$) and invokes the Material Rule Validator (encapsulating Pymatgen~\citep{ong2013python}, ASE~\citep{hjorth2017atomic}, and RDKit~\citep{rdkit}) to conduct strict legality checks on stoichiometry and crystal structure, filtering out physically invalid candidates. Subsequently, for feasible candidates that still show large gaps to target objectives, the Interpretable Inference Engine (based on SHAP analysis~\citep{lundberg2017unified}) quantifies feature contributions to identify key constraints. The analysis is synthesized into an attribution diagnostic report $\mathcal{R}$, containing hard-constraint validity markers, per-objective soft-constraint gaps, root-cause diagnoses, and next-iteration strategy recommendations. Based on these diagnostic outcomes, the Meta Agent refines its hypothesis and adapts the strategy for the next iteration.

\subsection{Wet-Lab Procedure}\label{sec-method-wetlab}

To ensure the physical viability of the material recipes and process parameters generated by PeroMAS, we collaborate with a professional experimental team in the perovskite field to establish a standardized fabrication workflow for perovskite solar cells (PSCs). This workflow employs a classic p-i-n inverted architecture~\citep{jeng2014nickel}, effectively translating abstract AI design proposals into physical devices. The specific procedures are provided in Appendix~\ref{app:wetlab}.

\begin{table*}[!t]
    \centering
    \caption{
        \textbf{Comparative evaluation of backbone models on core agent modules.} 
        Metrics include tool invocation accuracy (Any/Strict order) and output validity (0--10).
        \textbf{Bold} indicates the best result, and \underline{underlined} denotes the second best.
    }
    \label{tab:atomic_breakdown}
    \setlength{\tabcolsep}{8.5pt} 
    \renewcommand{\arraystretch}{1.25} 
    \vspace{-5px}
    \begin{tabular}{ll l cccc}
        \hline
        \multirow{2}{*}{\textbf{Agent Module}} & \multirow{2}{*}{\textbf{Core Capability}} & \multirow{2}{*}{\textbf{Metric}} & \multicolumn{4}{c}{\textbf{Backbone Model}} \\
        \cline{4-7}
         & & & \textbf{GPT-4o} & \textbf{Claude-4.5} & \textbf{Gemini-2.5} & \textbf{DeepSeek-V3} \\
        \hline
        \multirow{4}{*}{\textbf{Miner Agent}} & \multirow{4}{*}{\shortstack[l]{Knowledge\\Acquisition}} 
          & Tool-Any-Order & \textbf{100.0\%} & 92.5\% & \underline{95.0\%} & \underline{95.0\%} \\
          & & Tool-In-Order & \textbf{97.5\%} & 92.5\% & \underline{95.0\%} & 92.5\% \\
          & & Validity (LLM) & \textbf{7.64} & \underline{6.75} & 5.17 & 6.46 \\
          & & Validity (Expert) & \textbf{6.80} & 5.80 & 4.80 & \underline{6.30} \\
        \hline
        
        \multirow{4}{*}{\textbf{Designer Agent}} & \multirow{4}{*}{\shortstack[l]{Composition \&\\Recipe Design}} 
          & Tool-Any-Order & 76.6\% & \textbf{93.3\%} & 83.3\% & \underline{86.6\%} \\
          & & Tool-In-Order & 73.3\% & 70.0\% & \underline{80.0\%} & \textbf{83.3\%} \\
          & & Validity (LLM) & 6.74 & \textbf{7.52} & 5.62 & \underline{6.85} \\
          & & Validity (Expert) & 6.40 & \textbf{7.50} & \underline{6.16} & 5.50 \\
        \hline
        
        \multirow{4}{*}{\textbf{Emulator Agent}} & \multirow{4}{*}{\shortstack[l]{Performance\\Predictive}} 
          & Tool-Any-Order & \underline{94.2\%} & 88.5\% & 90.2\% & \textbf{100.0\%} \\
          & & Tool-In-Order & \underline{91.4\%} & 85.7\% & 88.5\% & \textbf{97.1\%} \\
          & & Validity (LLM) & \textbf{8.62} & 8.26 & 8.05 & \underline{8.33} \\
          & & Validity (Expert) & \textbf{7.33} & \underline{7.11} & 6.56 & 6.56 \\
        \hline
        
        \multirow{4}{*}{\textbf{Analyst Agent}} & \multirow{4}{*}{\shortstack[l]{Visual Analytics\\\& Interpretability}} 
          & Tool-Any-Order & 90.0\% & 90.0\% & 90.0\% & \textbf{95.0\%} \\
          & & Tool-In-Order & 75.0\% & \textbf{90.0\%} & 80.0\% & \underline{85.0\%} \\
          & & Validity (LLM) & \underline{8.15} & 7.76 & \textbf{8.26} & 7.65 \\
          & & Validity (Expert) & \textbf{8.00} & \underline{7.87} & 7.62 & \underline{7.87} \\
        \hline
        \addlinespace[0.2em]
        
        \multirow{2}{*}{\textbf{Overall}} & \multirow{2}{*}{System Average} 
          & \textbf{Avg. Tool Acc.} & 84.3\% & 84.5\% & \underline{85.9\%} & \textbf{89.5\%} \\
          & & \textbf{Avg. Validity} & \textbf{7.13} & \underline{7.07} & 6.28 & 6.56 \\
        \hline
    \end{tabular}
\end{table*}

\section{Experiments}\label{sec-experiments}

\subsection{Experimental Setup}

\subsubsection{Training Datasets And Tool Calls}

We construct a full-chain MCP (Model Context Protocol) toolchain driven by real experimental data. We select The Perovskite Database~\citep{jacobsson2022open} and execute a rigorous cleaning and completion process to obtain five-thousand pieces of full-dimensional ``component-process-performance'' data of perovskite in real scenarios. This data set forms the core knowledge base of the system and is directly used to train and fine-tune various specialized MCP tool models called by the agent.

\subsubsection{Evaluation Task Setup}
\label{evaluation_setup}
We designed a hierarchical evaluation protocol bridging "atomic execution" and ``system-level decision-making'' to comprehensively assess PeroMAS. This protocol will test the system from two dimensions: ``single point task execution'' to ``full-process closed-loop decision-making'':

\textbf{Atomic Capability Evaluation.}
We construct 130 specific tasks to strictly verify the execution accuracy of each individual agent.
\begin{itemize}[leftmargin=*]
    \item Miner Task: Evaluates the precision of literature retrieval and structured parameter extraction (composition, process, performance).
    \item Designer Task: Assesses the validity of inverse design proposals and the accuracy of physical feasibility judgments under hard constraints.
    \item Emulator Task: Tests the parallel processing capability for predicting key properties, specifically Power Conversion Efficiency (PCE) and Stability ($T_{80}$).
    \item Analyst Task: Verifies the logic of feature attribution analysis and the scientific soundness of mechanism deduction reports.
\end{itemize}

\textbf{System-Level Evaluation}
To evaluate collaborative problem-solv\-ing, we design 20 exploration tasks under conflicting multi-objective constraints (e.g., efficiency vs. stability vs. toxicity). This phase focuses on assessing the Meta Agent's dynamic orchestration strategy and overall scientific rationality of the final experimental solutions.

\subsubsection{Baselines}

To benchmark our PeroMAS, we compare it against three representative paradigms ranging from tool-free inference to generalist autonomous agents:

\begin{description}[style=unboxed, leftmargin=0cm, font=\bfseries]

    \item[Standard LLM:] 
    Uses the general-purpose \texttt{gpt-4o} for direct question-answering without access to any external MCP tools or domain knowledge bases. 

    \item[Single ReAct Agent:] 
    A monolithic agent (e.g. GPT-Agent) built on LangChain/AutoGPT~\citep{yang2023auto} architectures. While equipped with MCP tool access, it processes tasks via a single linear ReAct loop, lacking the hierarchical planning capabilities of our system.

    \item[Generalist Autonomous Agent:] 
    The generalist frameworks like Manus~\citep{shen2025mind}. We deploy it in the same perovskite environment to highlight the necessity of domain-specific architectural design over general-purpose autonomy.

\end{description}

\subsubsection{Evaluation Metrics}

To better quantify the results of the hierarchical evaluation protocol, we adopt two independent metric systems corresponding to the atomic capability evaluation of single agents and the full-loop evaluation of PeroMAS.

\textbf{Single-Agent Metrics}
We assess atomic precision using three indicators:
\textbf{Tool Accuracy (\%)} evaluates execution correctness across two dimensions: Any-Order verifies the selection of the correct tool set, while Strict-Order enforces logical sequential execution consistent with the workflow.
\textbf{Validity (0--10)} quantifies the scientific rationality of outputs, averaged from two sources: LLM-Score (assessing logical coherence via an independent evaluator) and Expert-Score (blinded peer review focusing on scientific value).

\textbf{System-Level Metrics}
For complex multi-objective scenarios,
\textbf{Task Completion (\%)} measures the success rate of generating recipes that pass hard-constraint validation and iteratively refine toward soft-constraint targets.
\textbf{Output Validity (0--10)} evaluates the comprehensive quality of the final solution. Similar to single-agent metrics, this integrates automated assessments of report logic with professional expert evaluations regarding physical feasibility, constraint-gap diagnosis, and potential impact.

\begin{figure}[!t]
  \centering
  \begin{subfigure}[t]{0.48\linewidth}
    \centering
    \includegraphics[width=\linewidth]{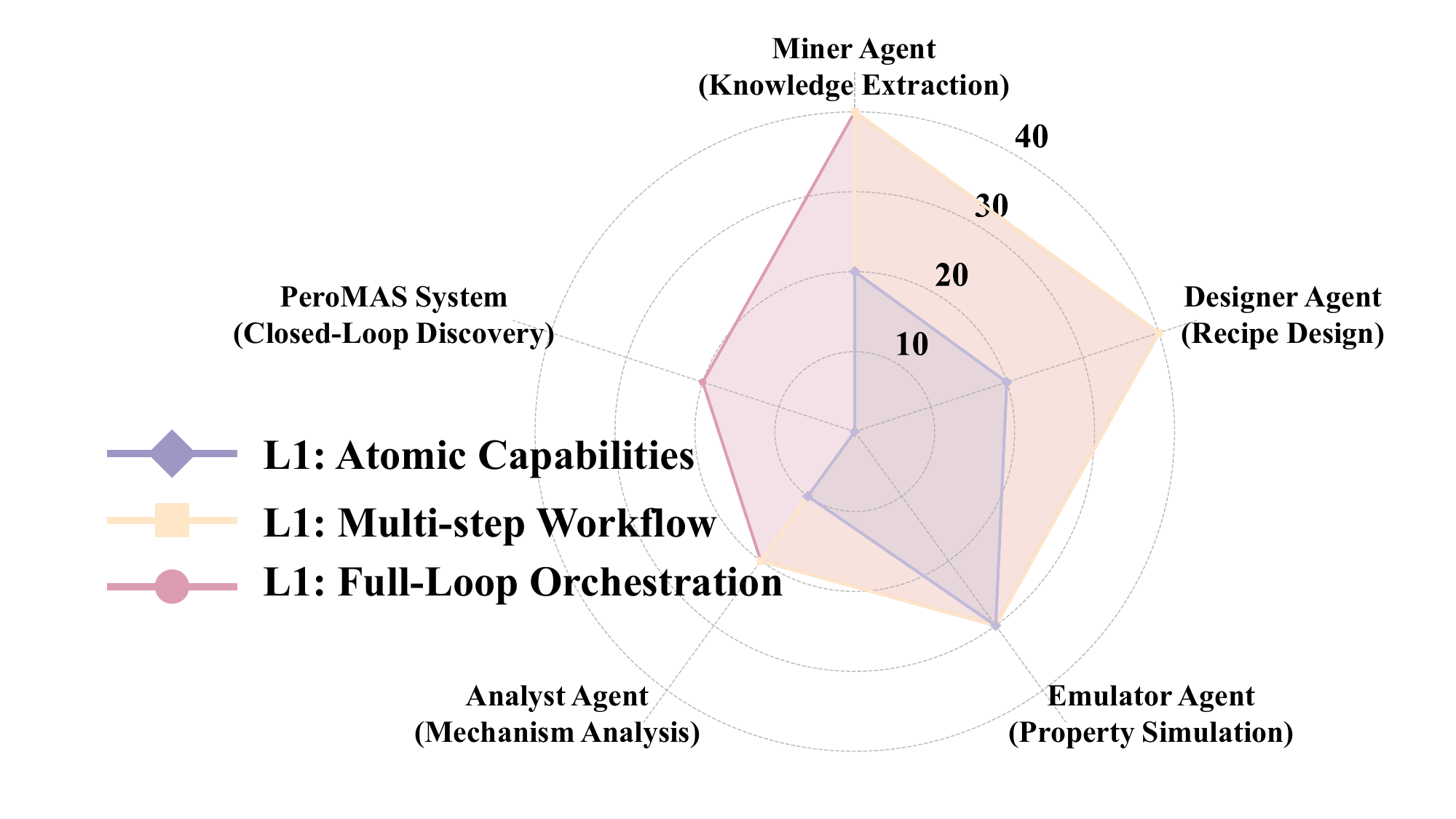}
    \caption{} 
    \label{fig:benchmark_complexity_levels}
  \end{subfigure}
  \hspace{0.02\linewidth}
  \begin{subfigure}[t]{0.48\linewidth}
    \centering
    \includegraphics[width=\linewidth]{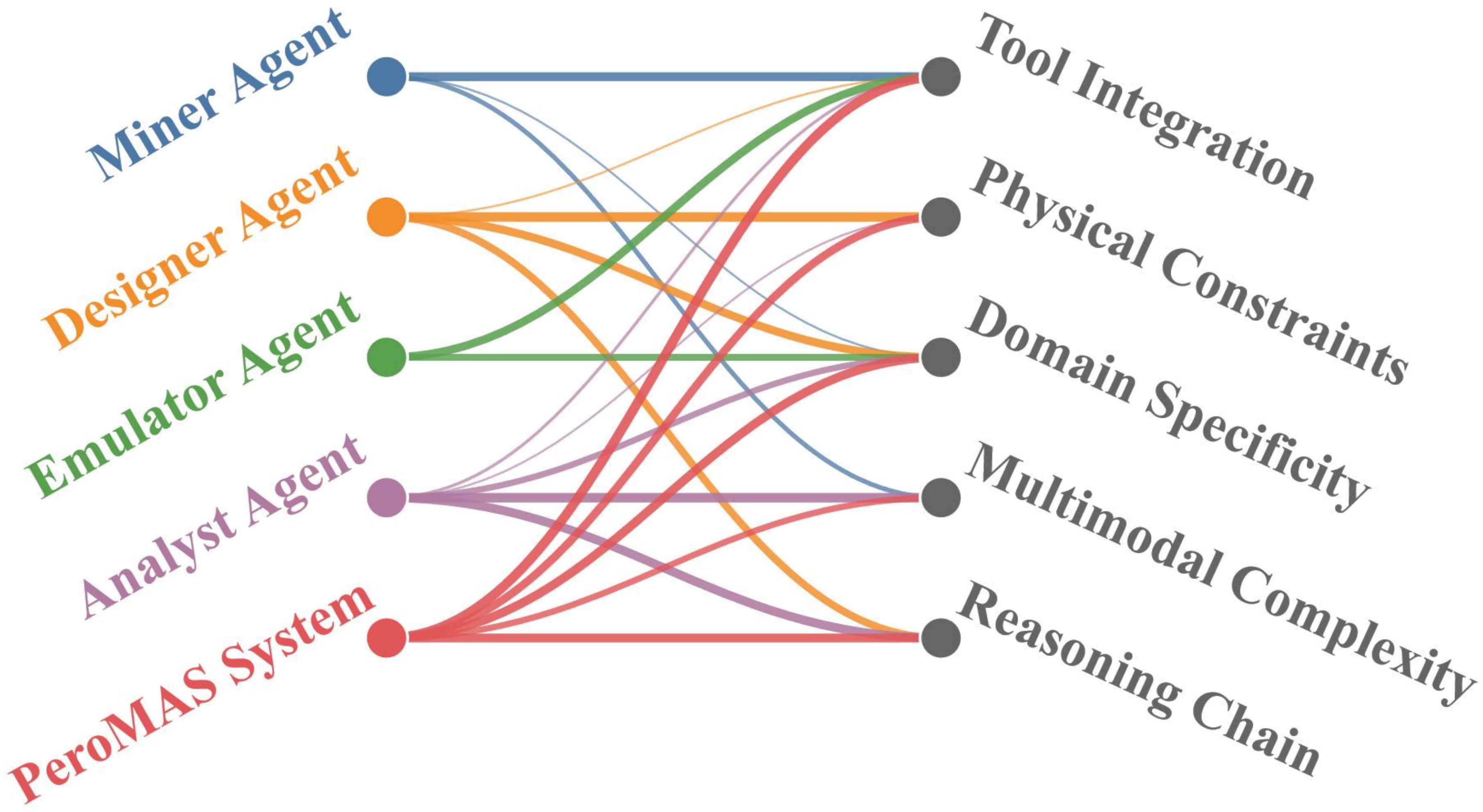}
    \caption{} 
    \label{fig:benchmark_quality_flow}
  \end{subfigure}
  \vspace{-10pt}
  \caption{\textbf{Overview of PeroMAS-Bench dataset.} 
  \textbf{(a)} Hierarchical task distribution across three complexity levels. 
  \textbf{(b)} Capability contribution flow mapping agents to qualitative dimensions, where thickness indicates challenge intensity.}
  \label{fig:benchmark_overview}
  \vspace{-12pt}
\end{figure}

\subsubsection{PeroMAS Benchmark}
We introduce PeroMAS-Bench, the first hierarchical evaluation dataset specifically designed for the perovskite domain. Constructed via the rigorous annotation pipeline detailed in Section~\ref{evaluation_setup}, the benchmark comprises 150 high-fidelity evaluation instances that ensure comprehensive coverage of the research lifecycle. As shown in Figure~\ref{fig:benchmark_complexity_levels}, the multi-layered architecture spans from foundational Miner tasks ($N=40$) for literature extraction to closed-loop discovery by PeroMAS ($N=20$).

To quantify dataset quality, we establish five cognitive dimensions: Tool Integration, Physical Constraints, Domain Specificity, Multimodal Complexity, and Reasoning Chain. 
As visualized in Figure~\ref{fig:benchmark_quality_flow}, the task-dimension binding is determined by the dominant capability required for valid execution: Miner tasks are rooted in Tool Integration for external retrieval, Designer tasks are bound by Physical Constraints to guarantee synthesizability, whereas Analyst tasks necessitate Multimodal Reasoning to interpret spectral data. All ground truth labels were verified by domain experts.

\subsubsection{Implementation Details}

\begin{description}[style=unboxed, leftmargin=0cm, font=\bfseries]

    \item[Backbone Models:] 
    We benchmark PeroMAS across representative closed-source (GPT-4o, Claude-4.5, Gemini-2.5) and open-source (DeepSeek-V3, Qwen3) architectures. In each experimental group, all agents are uniformly instantiated with the same backbone model to ensure fair comparison.

    \item[Hyperparameter Tuning:] 
    Generation parameters are tailored to specific agent roles to balance precision and creativity:
    \begin{itemize}[leftmargin=*]
        \item Miner \& Analyst ($\tau = 0.1$): Prioritizes determinism to ensure precise data extraction and rigorous attribution logic.
        \item Designer ($\tau = 0.7$, Top-$p=0.9$): Increases randomness to stimulate exploration within the vast material solution space.
        \item Meta Agent ($\tau = 0.3$): Balances stability in global orchestration with flexibility in strategy adjustment.
    \end{itemize}

    \item[Automated Evaluation Setup:] 
    We employ GPT-5.2 as the independent evaluator for validity metrics. To minimize variance, the evaluator temperature is locked at $\tau = 0.1$, following a strict double-blind protocol to assess both the planning trajectory and the final output.

\end{description}

\subsubsection{Wet-Lab Evaluation}

\begin{figure*}[!t]
  \centering
  \begin{subfigure}[t]{0.48\linewidth}
    \centering
    \includegraphics[width=\linewidth,height=4.0cm]{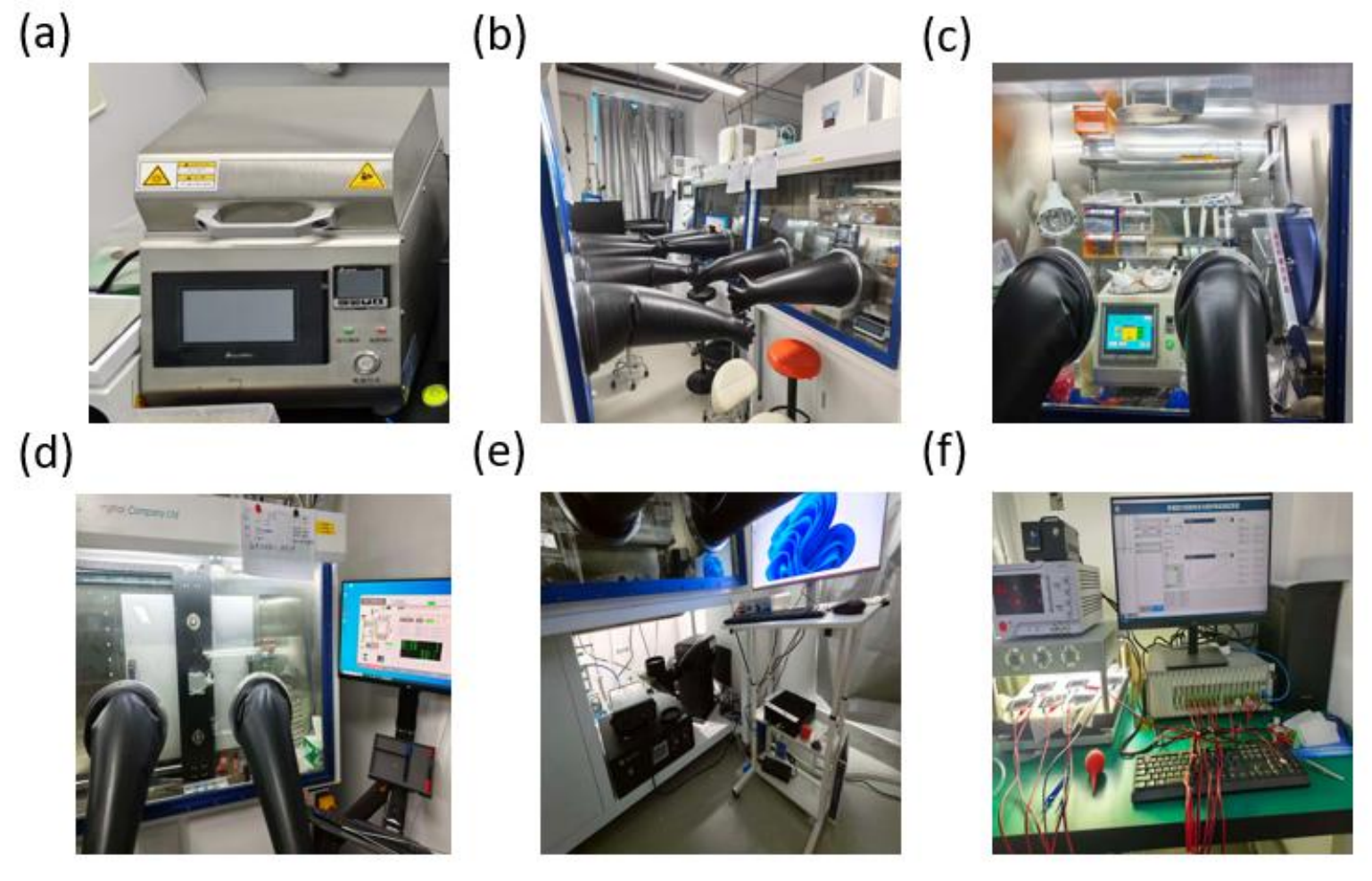}
    \caption{\textbf{Wet-Lab Procedure}} 
    \label{fig:wet_lab_setup}
  \end{subfigure}
  \hfill
  \begin{subfigure}[t]{0.48\linewidth}
    \centering
    \includegraphics[width=\linewidth,height=4.0cm]{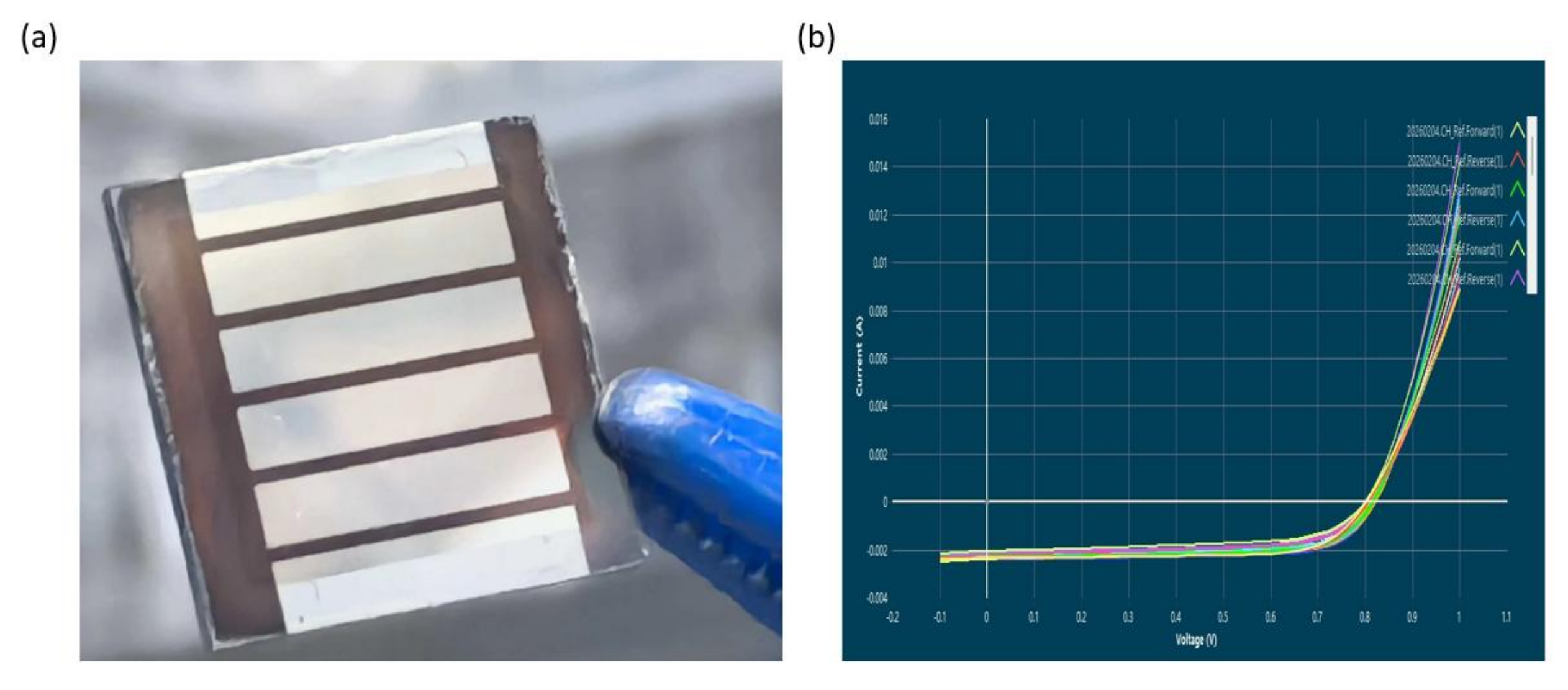} 
    \caption{\textbf{Wet-Lab Results}}
    \label{fig:wet-lab-result}
  \end{subfigure}
  \vspace{-5pt}
  \caption{\textbf{Practical validation of the PeroMAS system.} 
  \textbf{(a)} Used wet-lab devices for perovskite material synthesis. 
  \textbf{(b)} Experimental results displaying the fabricated perovskite devices and their corresponding J-V curves for PCE evaluation.}
  \label{fig:wet_lab_full}
  \vspace{-5pt} 
\end{figure*}

To evaluate the real-world feasibility of PeroMAS, we establish a constrained optimization scenario under efficiency-oriented and low-toxicity constraints. Our experiments are conducted in a standardized perovskite device fabrication laboratory equipped with a full-chain fabrication and characterization platform, as shown in Figure~\ref{fig:wet_lab_setup}. All precursor preparation and thin-film deposition processes are carried out within nitrogen-filled gloveboxes to maintain a strictly inert atmosphere. We utilize a precision spin coater to execute the AI-generated spin-coating recipes, followed by electrode deposition via a high-vacuum thermal evaporation system. The final device performance is characterized under a standard AM 1.5G solar simulator. This controlled experimental environment supports the translation of AI-generated strategies into physical validation.

\subsection{Results And Analysis}

We will present a comprehensive analysis of the experimental results across five key dimensions: atomic performance, system-level performance, evaluation reliability, comparative benchmarking, and wet-lab validation of the generated proposals.

\subsubsection{Atomic Performance}

Table ~\ref{tab:atomic_breakdown} highlights significant performance divergence across models. DeepSeek-V3 achieves a remarkable 89.5$\%$ tool accuracy, proving that open-source models can achieve functional parity with closed-source baselines in complex execution tasks. Claude-4.5 leads in overall output quality (Expert Score: 8.00) and complex recipe design, indicating that superior reasoning capabilities still demand the advanced cognitive depth of top-tier models. 

\subsubsection{System-Level Performance}

Table~\ref{tab:system_performance} demonstrates that Claude-4.5 establishes the SOTA, achieving both the highest task completion (72.4\%) and superior output validity (Expert Score: 7.14). 
Notably, the open-source DeepSeek-V3 outperforms GPT-5.2 specifically in Task Completion (61.8\% vs. 59.0\%). 
This confirms that while top-tier proprietary models dominate in scientific reasoning quality, well-orchestrated open-source agents can effectively bridge the execution gap in constrained tasks.

\begin{table}[!t]
    \centering
    \caption{
        \textbf{Performance evaluation of the fully autonomous perovskite material discovery workflow.} 
        \textbf{Bold} indicates the best performance, and \underline{underlined} indicates the second best.
    }
    \vspace{-10px}
    \label{tab:system_performance}
    \small
        \begin{tabular}{l c c c}
            \toprule
            \textbf{Base Model} & \textbf{ Completion} & \textbf{Validity (LLM)} & \textbf{Validity (Expert)} \\
            \midrule
            GPT-5.2 & 59.0\% & \underline{6.52} & 6.40 \\
            Gemini-2.5 & 48.7\% & 6.15 & 6.24 \\
            Claude-4.5 & \textbf{72.4\%} & \textbf{7.22} & \textbf{7.14} \\
            DeepSeek-V3 & \underline{61.8\%} & 6.35 & \underline{6.59} \\
            Qwen3 & 52.6\% & 6.05 & 5.92 \\
            \bottomrule
        \end{tabular}
        \vspace{-5px}
\end{table}

\subsubsection{Evaluation Consistency Analysis}

Figure~\ref{fig:evalusation_consistency} compares the score distributions between the LLM Judge and human experts. The two trajectories are highly congruent, with an average divergence of only $\pm$0.15 points across all models. This tight alignment confirms that, under rigorous scoring rubrics, GPT-5.2 model provides objective assessments highly consistent with expert judgment, thereby validating the reliability of our evaluation framework.

\subsubsection{Baseline Comparison Analysis}
\begin{figure}[!t]
  \centering
  \begin{subfigure}[t]{0.35\linewidth}
    \centering
    \includegraphics[width=\linewidth]{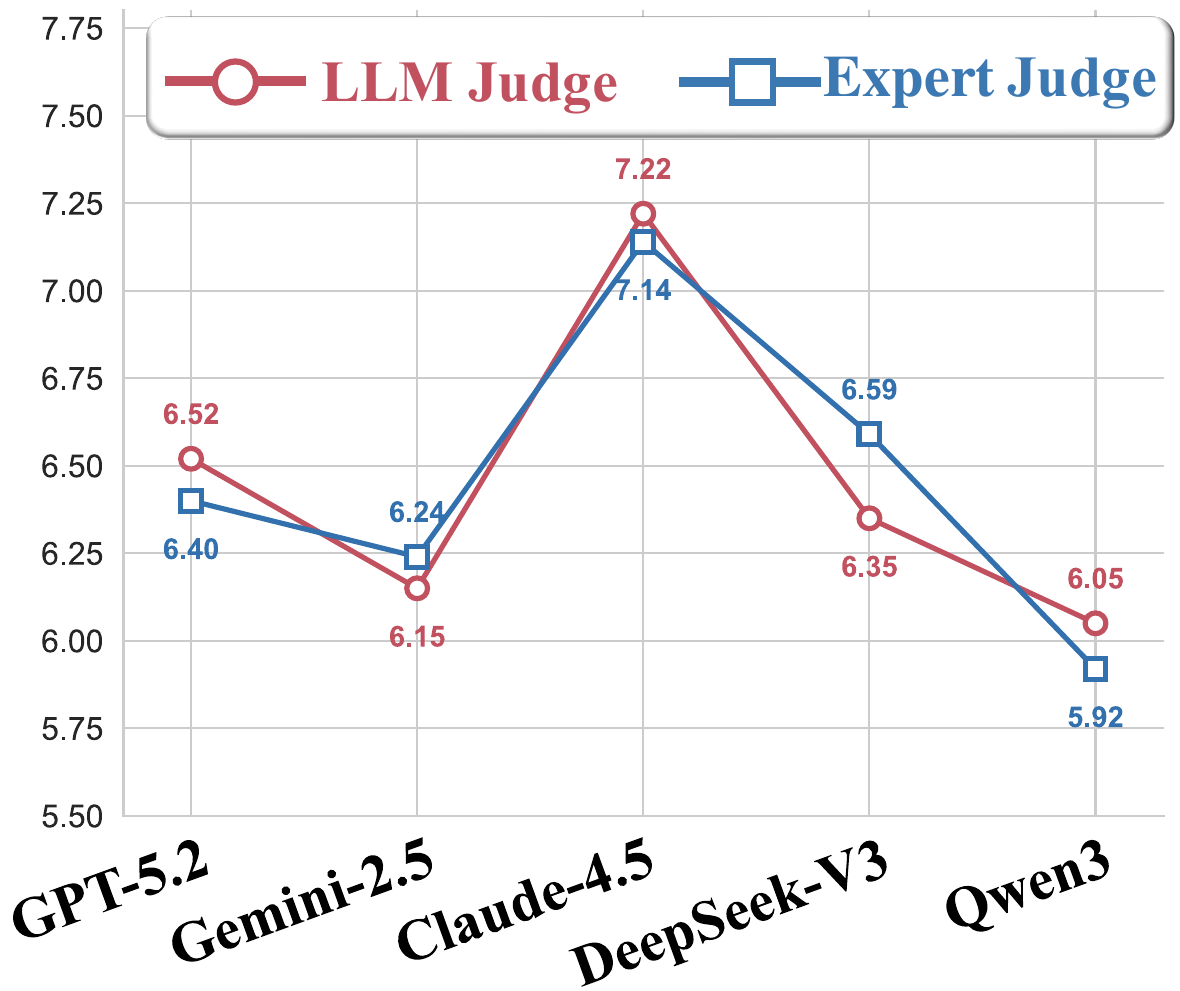}
    \caption{\textbf{Alignment}} 
    \label{fig:evalusation_consistency}
  \end{subfigure}
  \hfill 
  \begin{subfigure}[t]{0.55\linewidth}
    \centering
    \includegraphics[width=\linewidth]{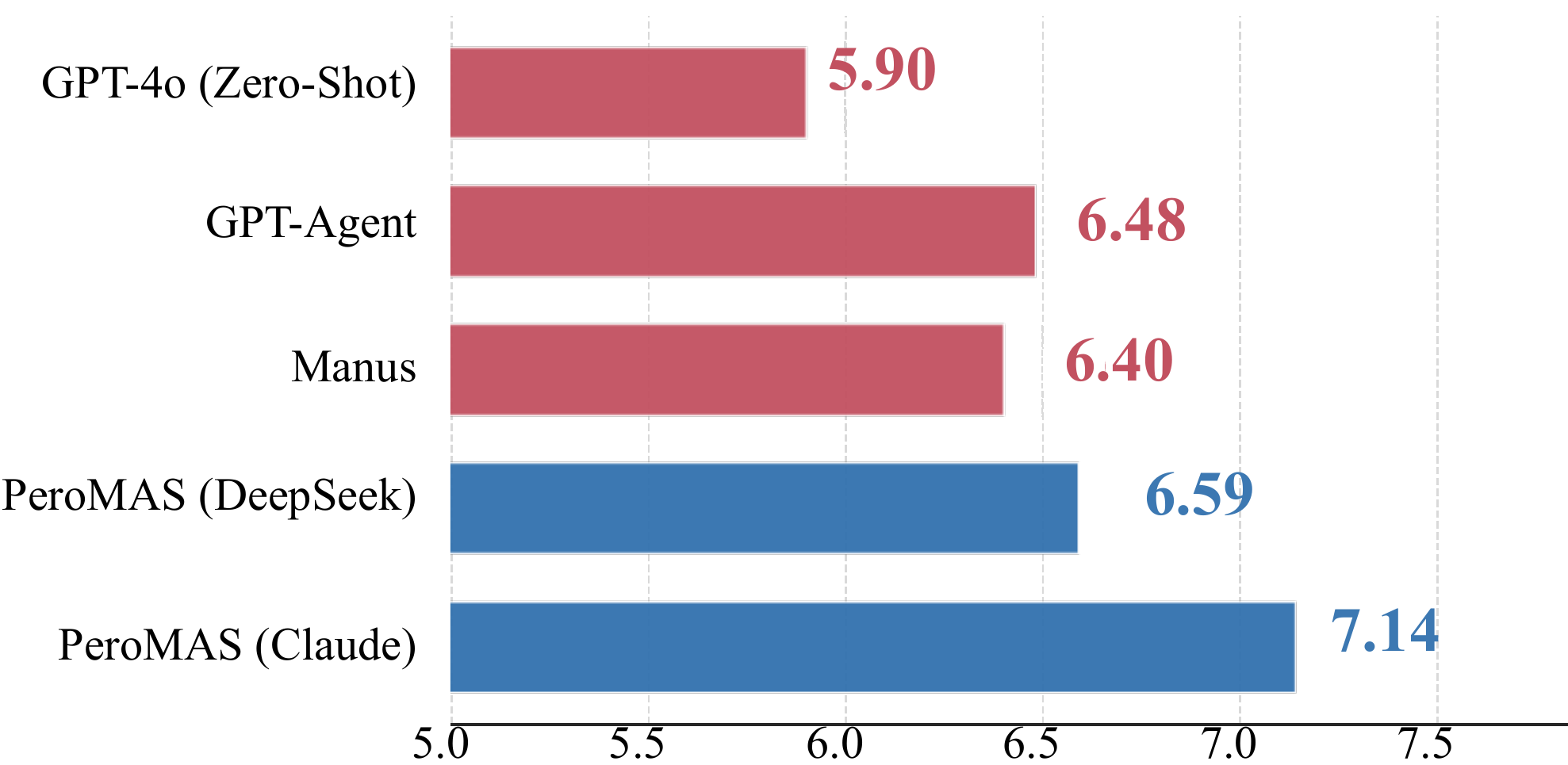}
    \caption{\textbf{Baseline Comparison}} 
    \label{fig:baseline_compare}
  \end{subfigure}
  
  \vspace{-5pt} 
  
  \caption{\textbf{Evaluation consistency and performance benchmarking.} 
  \textbf{(a)} Alignment of score distributions between LLM Judge and human experts. 
  \textbf{(b)} Performance comparison of PeroMAS against representative baselines across diverse agentic paradigms.}
  \label{fig:eval_analysis}
\end{figure}

To justify the necessity of a multi-agent architecture for perovskite discovery, we compare PeroMAS against three different paradigms, as shown in Figure~\ref{fig:baseline_compare}. The results yield three key architectural insights:

\begin{description}[style=unboxed, leftmargin=0cm, font=\bfseries]

    \item[Necessity of External Tools:] 
    The Standard LLM (GPT-4o) scores lowest (5.90), confirming that internal parametric knowledge is insufficient for complex discovery. Integrating external MCP toolchains is a fundamental prerequisite for advanced scientific research.

    \item[Superiority of Domain Adaptation:] 
    The generalist agent Manus underperforms the domain-specialized GPT-Agent. This reveals the limitations of "general-purpose autonomy" in niche scientific fields, which indicates that equipping agents with domain-specific tools yields greater value than generalist agents.

    \item[Collaborative Architecture Advantage:] 
    Using identical tools, PeroMAS achieves a 10.2\% performance gain over the single-agent baseline, validating that hierarchical task decomposition effectively reduces cognitive load. Notably, PeroMAS (DeepSeek-V3) outperforms proprietary baselines (Manus, GPT-Agent), suggesting that a superior collaborative framework outweighs raw model scale in multi-objective scientific challenges.

\end{description}

\subsubsection{Wet-Lab Results}
We select the Top-1 perovskite recipe generated by PeroMAS for physical synthesis and performance characterization. The selected recipe is a narrow-bandgap Sn--Pb mixed perovskite formulated as (FASnI$_3$)$_{0.6}$(MAPbI$_3$)$_{0.4}$, with 10\% SnF$_2$ as the reducing additive and 3\% EDAI$_2$ as the passivation additive. This composition reduces the Pb fraction by approximately 50\% compared with conventional pure-Pb perovskites while targeting a bandgap of about 1.23 eV. Following the system-provided fabrication protocol, we synthesize perovskite thin films and record a PCE of 17\% through J-V characteristic testing, as shown in Figure~\ref{fig:wet-lab-result}. This empirical performance falls within the predicted PCE range of 16\%--19\%, demonstrating consistency between dry-lab prediction and physical validation.

\section{Limitations and Ethical Considerations}

\noindent\textbf{Limitations.}
PeroMAS is currently limited by knowledge coverage, data scale, evaluation depth, and physical automation. First, incomplete literature coverage may cause the system to miss useful priors or better candidate solutions; future work will expand data sources and support incremental updates. Second, the current dataset covers only part of the PSC composition and processing space, which may affect property prediction and candidate ranking; integrating more public databases can improve robustness. Third, our analysis does not yet include systematic cross-backbone comparison or detailed attribution; future work will examine how different agents, tools, and models affect performance. Finally, the system has not been connected to batch wet-lab automation, and integrating automated synthesis and characterization platforms will be needed for a fully closed-loop discovery system.

\noindent\textbf{Ethical Considerations.}
From an ethical perspective, the system should be used as a decision-support tool rather than a replacement for domain experts. AI-generated experimental suggestions may contain incomplete, unsafe, or non-actionable procedures and should be reviewed by qualified researchers before wet-lab execution. In addition, PSC research may involve lead-containing materials, organic solvents, and chemical waste; therefore, all experiments should follow standard safety, toxicity, and waste-disposal protocols. Since the system relies on literature, databases, external tools, and LLMs, its outputs may inherit data bias, measurement noise, model hallucinations, or reliability risks in automated scientific judgment~\citep{wang2026ai}. We therefore recommend using it to support responsible PSC discovery under expert oversight.

\section{Conclusion}
We present PeroMAS, a hierarchical multi-agent framework for perovskite material discovery. By integrating MCP-based tools, PeroMAS mitigates the fragmentation of single-task models and supports iterative constraint tracking, diagnosis, and refinement across the discovery workflow. Evaluations show that PeroMAS outperforms generalist agents, with open-source backbones achieving competitive performance with proprietary models. Notably, wet-lab validation confirmed a 17\% PCE for an AI-designed Sn-Pb recipe, verifying the system's practical viability in iterative perovskite material optimization. The released PeroMAS Benchmark provides a foundation for evaluating PSC-oriented scientific agents and advancing future self-driving laboratories.

\section{Acknowledgments}

This work is funded by the Basic Research Program of Jiangsu under Grant Nos. BK20253021 and BK20251292, and the National Natural Science Foundation of China under Grant Nos. 62506075, 62225602, and 52503336. This research work is also supported by the Big Data Computing Center of Southeast University.

\clearpage



\bibliographystyle{ACM-Reference-Format}
\balance
\bibliography{refs}

\clearpage  
\appendix

\section{Toolchain Construction}
\label{app:Dataset}
\subsection{Training Dataset Preparation}

All datasets used in PeroMAS are derived from the open-access Perovskite Database Project~\cite{jacobsson2022open}, which contains 43,398 experimental PSC records across 410 attributes, covering composition, fabrication, device architecture, photovoltaic performance, and stability. We processed the raw database into task-specific sub-datasets for downstream modules, as summarized in Table~\ref{tab:dataset_overview}.

\textbf{Data Processing Pipeline.}
The preprocessing includes four main steps: (1) filtering entries with valid perovskite compositions and recognized B-site metals (e.g., Pb, Sn, Bi); (2) cleaning critical attributes such as solvent systems, deposition methods, and additives; (3) removing duplicates while preserving diverse process-property mappings; and (4) augmenting matched compositions with crystal structures and DFT-calculated properties from the Materials Project~\cite{jain2013commentary}, achieving a 75.2\% match rate.

\begin{table}[H]
    \centering
    \caption{\textbf{Overview of task-specific datasets constructed for PeroMAS.} The data is categorized by the functional module it supports.}
    \label{tab:dataset_overview}
    \small
    \setlength{\tabcolsep}{4pt}
    \renewcommand{\arraystretch}{1.05}
    \begin{tabularx}{\linewidth}{X c c}
    \toprule
    \textbf{Task / Functional Module} & \textbf{Split} & \textbf{Total} \\
    \midrule
    \multicolumn{3}{l}{\textbf{\textit{1. Synthesis Feasibility Assessment}}} \\
    \hspace{1em} Positive Samples & 2,608 / 290 & 2,898 \\
    \hspace{1em} Negative Samples & --- & 700 \\
    \midrule
    \multicolumn{3}{l}{\textbf{\textit{2. Synthesis Route Planning}}} \\
    \hspace{1em} Method Recommendation & 4,500 / 500 & 5,000 \\
    \hspace{1em} Precursor Identification & --- & 2,296 \\
    \midrule
    \multicolumn{3}{l}{\textbf{\textit{3. Crystal Structure Generation}}} \\
    \hspace{1em} Core Set & 639 / 72 & 711 \\
    \hspace{1em} Expanded Set & 37,296 / 4,145 & 41,441 \\
    \midrule
    \multicolumn{3}{l}{\textbf{\textit{4. Property Prediction}}} \\
    \hspace{1em} Multi-target Regression & $\sim$569 / 142 & 711 \\
    \bottomrule
    \end{tabularx}
\end{table}
Table~\ref{tab:source_columns} details the specific physical attributes extracted to support these tasks.

\begin{table}[H]
    \centering
    \caption{\textbf{Key attributes extracted from the Perovskite Database.}}
    \label{tab:source_columns}
    \small
    \begin{tabular}{l l}
    \toprule
    \textbf{Category} & \textbf{Attributes} \\
    \midrule
    Composition & A/B/X-site ions, stoichiometric coefficients \\
    Performance & PCE, $V_\mathrm{oc}$, $J_\mathrm{sc}$, Fill Factor (FF) \\
    Band gap & Experimental \& DFT-calculated values \\
    Fabrication & Deposition method, solvent system, annealing temp. \\
    Structure & CIF files (augmented via Materials Project) \\
    Stability & T80 lifetime, PCE retention rate \\
    Device & Device architecture (n-i-p / p-i-n), layer thickness \\
    \bottomrule
    \end{tabular}
\end{table}

\subsection{MCP Tool}
\label{app:mcp_specs}

To standardize the interaction between agents and external knowledge, we encapsulated specialized algorithms and data sources into the Model Context Protocol (MCP), following recent MCP-based tool-service efforts~\citep{ouyang2025code2mcp, di2026toolrosetta}. Table~\ref{tab:mcp_tools} details the core toolset assigned to each functional agent.

\begin{table}[H]
    \centering
    \caption{\textbf{List of specialized MCP tools employed by PeroMAS.} Citations indicate the underlying algorithms or data sources.}
    \label{tab:mcp_tools}
    \small
    \resizebox{\linewidth}{!}{
        \begin{tabular}{l l p{5cm}}
        \toprule
        \textbf{Target Agent} & \textbf{MCP Tool ID} & \textbf{Backend / Description} \\
        \midrule
        \multirow{3}{*}{\textbf{Miner}} 
         & \texttt{arxiv-mcp} & Real-time literature retrieval via arXiv API \\
         & \texttt{Sci-mcp} & Cross-database search (ScienceDirect, Springer) \\
         & \texttt{extract-mcp} & Hybrid information extraction engine integrating LLM~\cite{wan2024tokens} and SciBERT~\cite{beltagy2019scibert} \\
        \midrule
        \multirow{3}{*}{\textbf{Designer}} 
         & \texttt{MatterGen} & Generative model for crystal structure inverse design~\cite{zeni2025generative} \\
         & \texttt{CSLLM} & Synthesis feasibility reasoning and recipe recommendation~\cite{song2025accurate} \\
         & \texttt{MatProj-mcp} & Structure and property querying via Materials Project API~\cite{jain2013commentary} \\
        \midrule
        \multirow{2}{*}{\textbf{Emulator}} 
         & \texttt{CGCNN} & Graph neural network for bandgap and formation energy prediction~\cite{xie2018crystal} \\
         & \texttt{Pero-ML} & Data-driven regression models (XGBoost,RF,NN)~\cite{hu2024interpretable} for device performance ($PCE$, $V_{oc}$) \\
        \midrule
        \multirow{3}{*}{\textbf{Analyst}} 
         & \texttt{SHAP-mcp} & Interpretable machine learning analysis for feature attribution~\cite{lundberg2017unified} \\
         & \texttt{Pymatgen} & Geometric analysis and structure manipulation toolkit~\cite{ong2013python} \\
         & \texttt{ASE} & Atomic Simulation Environment for thermodynamic checks~\cite{hjorth2017atomic} \\
        \bottomrule
        \end{tabular}
    }
\end{table}

\section{Wet-Lab Fabrication Details}
\label{app:wetlab}

All physical validation experiments were conducted in a standardized ISO Class 7 cleanroom environment. The specific fabrication protocol for the p-i-n inverted perovskite solar cells includes three key stages:

\paragraph{\textbf{Substrate Preparation \& HTL}} 
Indium Tin Oxide (ITO) glass substrates ($15\,\Omega/\text{sq}$) were sequentially cleaned via ultrasonication in detergent, deionized water, acetone, and isopropanol for 20 minutes each. After drying, the substrates underwent UV-Ozone treatment for 15 minutes to improve wettability. A Nickel Oxide (NiO$_x$) nanoparticle dispersion was then spin-coated at 4000 rpm for 30s and annealed at 150$^\circ$C for 30 minutes in ambient air to form the hole transport layer.

\paragraph{\textbf{Perovskite Deposition}}
The substrates were transferred to a nitrogen-filled glovebox (O$_2 < 0.1$ ppm, H$_2$O $< 0.1$ ppm). The perovskite precursor solution, prepared according to the exact stoichiometry output by the Designer Agent, was spin-coated using a one-step method. The specific spin program typically involved 1000 rpm for 10s and 4000 rpm for 30s, with anti-solvent (chlorobenzene) dropped 10s prior to the end. The films were subsequently annealed at 100$^\circ$C for 10--60 minutes (depending on the agent's recipe) to promote crystallization.

\paragraph{\textbf{ETL \& Electrode Evaporation}}
Finally, the Electron Transport Layer (ETL) consisting of C$_{60}$ (20 nm) and Bathocuproine (BCP, 5 nm), followed by a Silver (Ag) electrode (100 nm), were sequentially deposited via thermal evaporation under a high vacuum of $5 \times 10^{-4}$ Pa. The active area of each device was defined as $0.09\,\text{cm}^2$ using a metal shadow mask.
\end{document}